\documentclass[english,aps,prl,twocolumn]{revtex4-1}
\usepackage{amsmath,amssymb}
\usepackage[T1]{fontenc}
\usepackage[latin9]{inputenc}
\usepackage{float}
\usepackage{graphicx}
\usepackage{esint}

\makeatletter
\@ifundefined{textcolor}{}
{%
 \definecolor{BLACK}{gray}{0}
 \definecolor{WHITE}{gray}{1}
 \definecolor{RED}{rgb}{1,0,0}
 \definecolor{GREEN}{rgb}{0,1,0}
 \definecolor{BLUE}{rgb}{0,0,1}
 \definecolor{CYAN}{cmyk}{1,0,0,0}
 \definecolor{MAGENTA}{cmyk}{0,1,0,0}
 \definecolor{YELLOW}{cmyk}{0,0,1,0}
 }
\newcommand{\ba}{\begin{eqnarray}}
\newcommand{\ea}{\end{eqnarray}}

\newcommand{\ignore}[1]{}
\newcommand{\pb}{{\textbf{p}}}

\makeatother

\usepackage{babel}
\begin{document}
\global\long\def\bra#1{\left\langle #1\right|}

\global\long\def\ket#1{\left|#1\right\rangle }

\global\long\def\bk#1#2#3{\bra{#1}#2\ket{#3}}

\global\long\def\ora#1{\overrightarrow{#1}}

\preprint{This line only printed with preprint option}

\title{Topological Insulator Magnetic Tunnel Junctions: Quantum Hall Effect and Fractional Charge via Folding}

\author{Qinglei Meng, Smitha Vishveshwara, and Taylor L. Hughes}
\affiliation{Department of Physics, University of Illinois, 1110 West Green St, Urbana IL 61801}

\date{\today}
\begin{abstract}
We provide a characterization of tunneling between coupled topological insulators in 2D and 3D under the influence of a ferromagnetic layer. We explore conditions for such systems to exhibit integer quantum Hall physics and localized fractional charge, also taking into account interaction effects for the 2D case. We show that the effects of tunneling are topologically equivalent to a certain deformation or folding of the sample geometry. Our key advance is the realization that the quantum Hall or fractional charge physics can appear in the presence of only a \emph{single} magnet unlike previous proposals which involve magnetic domain walls on the surface or edges of topological insulators respectively. We give illustrative topological folding arguments to prove our results and show that for the 2D case our results are robust even in the presence of interactions. 
\end{abstract}
\maketitle
Time-reversal invariant topological insulators in 2D\cite{Kane2005a,Kane2005b,bernevig2006a,bernevig2006c,koenig2007} and 3D\cite{fu2007b,moore2007,roy2009a,hasan2010} serve as a novel platform for heterostructure devices. Because of the unique nature of the gapless boundary states in these nominally insulating materials, proximity coupling to ferromagnets\cite{qi2008} and s-wave superconductors\cite{fu2008} leads to exciting phenomena including fractional charge\cite{qi2008}, quantum Hall effect\cite{fu2007a,qi2008b}, topological magneto-electric effect\cite{qi2008b}, Majorana fermion bound states\cite{fu2008}, and exotic Josephson effects\cite{fu2009,akhmerov2009}. The origin of almost all of these effects is a consequence of the Dirac nature of the edge/surface fermions coupled to mass-inducing perturbations (\emph{e.g.} ferromagnets and s-wave superconductors) which are inhomogeneous. These proposals are popular because of their inherent practicality since they involve mature techniques with the only new ingredient being the topological insulator (TI).  For example, to produce a quantum Hall effect on the surface of a 3D TI, or analogously fractional charge on the edge of a 2D TI one need only deposit insulating ferro-magnets on the boundaries of these materials and induce a magnetic domain wall. The magnetic domain wall creates a mass-domain wall as seen by the gapless Dirac boundary states and all mass domain-walls trap low-energy bound states\cite{Jackiw1976}. It is these  boundstates which are responsible for the quantum Hall effect and fractional charge on the boundary of the 3D and 2D systems respectively. 

In this work we explore heterostructures which are made using two topological insulators with some non-zero tunneling processes between them. These heterostructures can be made using layered growth techniques (for 3D) and possibly even through lithography and gate patterning (for 2D). We provide a classification of conventional tunneling terms and indicate the interplay between tunneling, and proximity induced ferromagnetism and superconductivity. Interestingly, we find a new way to generate the integer quantum Hall effect (fractional charge) by utilizing only a \emph{single} ferromagnet sandwiched between two 3D (2D) TI's and without a magnetic domain wall. The proposed geometries are far simpler than those proposed in \cite{qi2008b,qi2008} which require the tuning of several magnetic regions. Along with the mathematical analysis we provide an intuitive, topological understanding of these effects.

We first focus on the case of the 3D TI and note, where important, the differences between 3D and 2D. The surface state Hamiltonian for a 3D TI is simply given by
\begin{eqnarray}
H=v\left(\boldsymbol\sigma\times \pb\right)\cdot \hat{{\textbf{n}}}
\end{eqnarray}\noindent where $\boldsymbol \sigma$ are the spin-1/2 matrices, $\pb$ is the surface momentum, and $\hat{\textbf{n}}$ is the normal vector to the relevant surface. Here we envision having two nearby 3D TI's separated by a distance $d_z$ in the z-direction. The decoupled, low-energy Hamiltonian for the bottom surface of the top TI and the top surface of the bottom TI is 
\begin{eqnarray}
\cal{H}&=&v\left(\begin{array}{cc} p_x\sigma_y-p_y\sigma_x & 0  \\ 0 & -p_x\sigma_y+p_y\sigma_x\end{array}\right)\\
&=&v\tau_z\otimes (p_x\sigma_y-p_y\sigma_x)
\end{eqnarray}\noindent where $\tau^a$ represent the layer index for the fermion basis $(\psi_{t\uparrow}\;\; \psi_{t\downarrow}\;\;\psi_{b\uparrow}\;\;\psi_{b\downarrow})^{T}$ where $t,b$ represent the top and bottom TI surfaces respectively. If the surfaces are coupled through time-reversal ($T=\mathbb{I}\otimes i\sigma_y K$) invariant tunneling processes that are also spin independent, which is natural from the form of the bulk Hamiltonian, then the only tunneling term is $H_t= t_{R}\tau^x\otimes\mathbb{I}.$ Since this matrix anti-commutes with ${\cal{H}}$ the spectrum is simply $E_{\pm}=\sqrt{p_{x}^2+p_{y}^2+t_{R}^2}$ (we have set $v=1$), which is gapped for all non-zero values of $t_R.$ To find all allowed mass terms in the double-layer system, we tabulate all $4\times 4$ matrices which anti-commute with $\cal{H}.$ There are four allowed terms which can open a gap. The first two are $m^{++}\mathbb{I}\otimes \sigma_z$ and $m^{+-}\tau_z\otimes\sigma_z$ and are non-zero if there is a magnetization with a component parallel to the surface normal which points in the same ($m^{++}\neq 0$) or opposite ($m^{+-}\neq 0$) direction in the two layers. The other two terms are the real and imaginary hopping terms $t_{R}\tau_x\otimes\mathbb{I}$ and $ t_{I}\tau_y\otimes \mathbb{I}.$ They all break time-reversal symmetry except $t_{R}.$ Similar terms have also been discussed in the context of a mean-field description of a topological exciton insulator\cite{seradjeh2009} although our different physical motivation is the key to our measurable predictions. 

For completeness, we also consider the possibility of proximity coupling to a superconductor to induce mass-terms\cite{fu2008}. The BdG equation in the Nambu spinor basis $(\psi_{t\uparrow},\psi_{t\downarrow},\psi_{b\uparrow},\psi_{b\downarrow},\psi_{t\downarrow}^{\dagger},-\psi_{t\uparrow}^{\dagger},\psi_{b\downarrow}^{\dagger},-\psi_{b\uparrow}^{\dagger})^{T}$
of this double-layer system is $ \mathcal{H}_{K}=\frac{1}{2}\pi_{z}\otimes\tau_{z}\otimes(p_{x}\sigma_{y}-p_{y}\sigma_{x}),$
where the Pauli matrices $\pi_{i}$ represent particle-hole space. Along with the magnetic and tunneling terms introduced above, surface gaps can be opened by inducing s-wave superconductivity with the same ($\Delta^{++}_{R/I}=\pi_{x/y}\otimes I\otimes I$) or opposite ($\Delta^{+-}_{R/I}=\pi_{x/y}\otimes \tau_z\otimes I$) phase on the two different TIs,  or even inter-TI s-wave pairing of Cooper pairs formed between the two TIs ($\Delta^{tb}_{R/I}=\pi_{y/x}\otimes \tau_y\otimes \sigma_z $).  Each of these possible pairing terms has both real and imaginary parts indicated by the $\pi_{x/y}$ notation giving six additional mass terms, yielding a total of ten. These ten mass terms can be arranged into a $3\times 3$ table and a $1\times 1$ table shown in Table \ref{tab:1}. This table has the following noteworthy properties: (i) the terms in all vertical or horizontal lines mutually \emph{commute} (ii) the terms in each diagonal line (including wrapping as if the table had periodic boundary conditions) mutually \emph{anti-commute} (iii)the left-over mass term $m^{++}$ \emph{commutes} with \emph{all} the other nine mass terms. 

\begin{table}[t]
\noindent \centering{}%
\begin{tabular}{|c|c|c|}
\hline 
$t_{R}$ & $\Delta_{R}^{+-}$ & $\Delta_{R}^{tb}$\tabularnewline
\hline 
$\Delta_{I}^{+-}$ & $m^{+-}$ & $\Delta_{I}^{++}$\tabularnewline
\hline 
$\Delta_{I}^{tb}$ & $\Delta_{R}^{++}$ & $t_{I}$\tabularnewline
\hline 
\end{tabular}\caption{\label{tab:1}This table of mass terms represents perturbations which open a gap in the surface-state Hamiltonian of two, coupled TI layers.  The three mass
terms in each vertical or horizontal line mutually commute with each other,
and  the three mass terms in each diagonal direction (including periodic wrapping) mutually anti-commute with each other. The remaining mass term $m^{++}$ commutes with \emph{all} of these nine mass
terms in the grid. }
\end{table}

This organization of mass terms is useful in considering the natural possibility of having more than one type of mass term present.  They can be classified pairwise as  (i) compatible masses and (ii) competing masses. Case (i) results from the two mass terms \emph{anti-commuting}. In this case, the energy spectrum takes the form $E_{\pm}=\pm\sqrt{p^2+m_{1}^{2}+m_{2}^2}$, from which it is easy to see that one can adiabatically tune between the phases dominated by $m_1$ ($m_2=0$) and the phase dominated by $m_2$ ($m_1=0$). This indicates that these two gapped phases are adiabatically connected. The competing mass case (ii) arises when the two mass terms \emph{commute}. In this case the spectrum generically takes the form $\pm E_{\pm}=\pm\sqrt{p^2 +(m_1\pm m_2)^2}.$ Thus, when going from the phase with $m_2=0$ to the phase with $m_1=0$ one always passes through a gapless critical point when the magnitudes of the mass terms are equal. Interestingly, if one places a region dominated by $m_1$ adjacent to a region dominated by $m_2$, there is a mass-domain wall between these regions that traps a low-energy fermionic bound state. This is the origin, for example, of the 0D Majorana fermion bound states at the interface between a magnet and superconductor placed on the edge of 2D TI\cite{fu2008}.  We can easily read-off the sets of compatible and competing mass-terms from Table \ref{tab:1}. 

\begin{figure}[t]
    \begin{center}
        \includegraphics[width=3.5in]{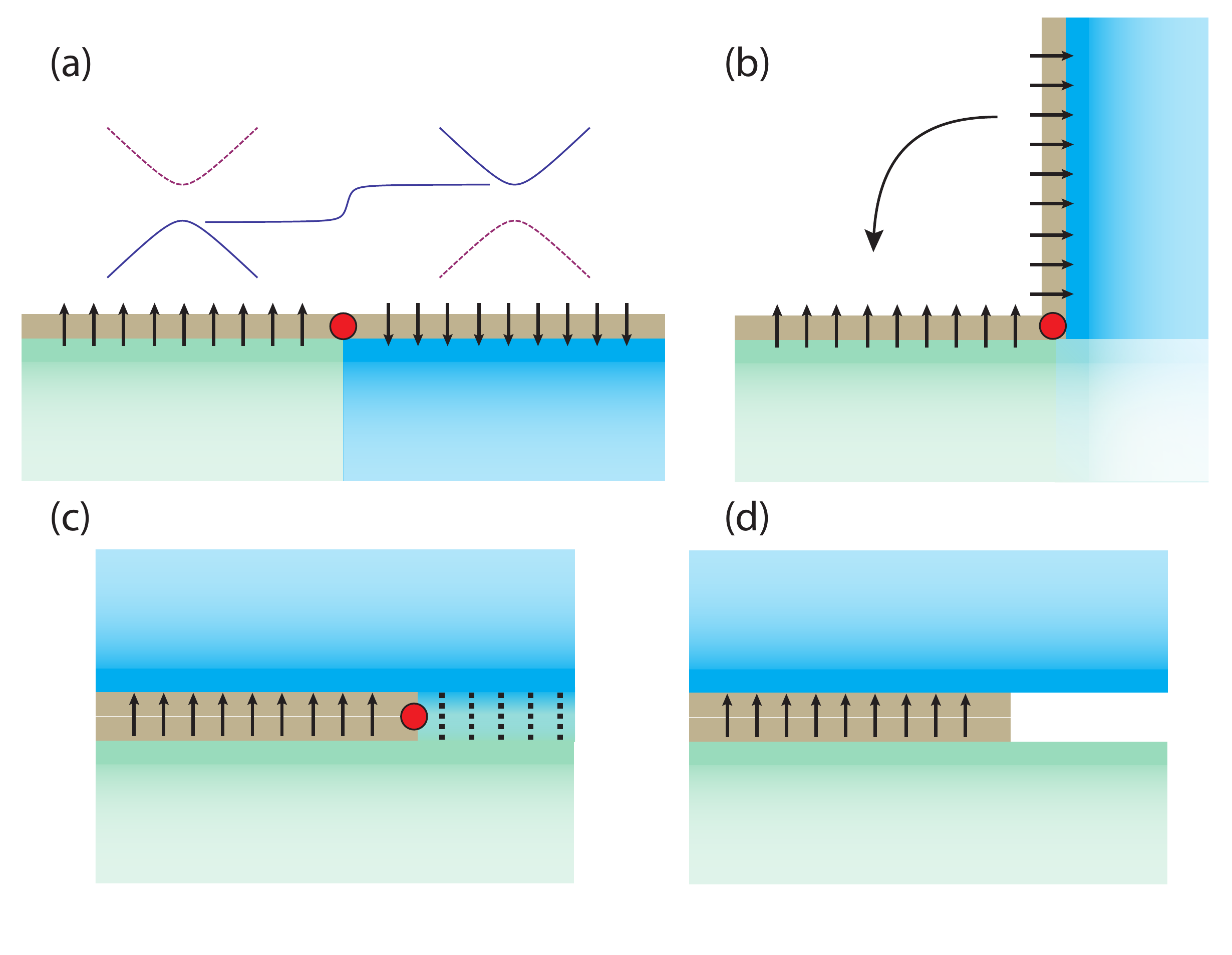}
    \end{center}
    \caption{ (a)Standard magnetic domain wall picture on the surface of a 3D TI (or edge of a 2D TI) which gives rise to a propagating chiral mode (bound $e/2$ charge). Inset: illustrates the mass domain wall seen by the Dirac fermions on the surface (edge) (b)An illustration of the folding process. Note that the chiral modes ($e/2$ charge) is preserved for every step in the fold. (c) Completion of folding/gluing process which shows that the final state is a domain wall between a ferromagnet and a tunneling region. The tunneling region is topologically equivalent to the bulk of the TI (d)When tunneling region is removed, leaving empty vacuum, the chiral modes ($e/2$ charge) are destroyed. }
    \label{fig:3dfolding}
\end{figure}

We use these results to study our primary case of interest: the tunneling $t_R$  competing with the magnetization $m^{++}$ (shortened to $m$ for convenience.) Suppose we place a magnet sandwiched between the two tunnel-coupled 3D topological insulators. Let the magnet have some component of its magnetization parallel to the z-direction so that a gap $m$ is induced in ${\cal{H}}.$ If $t_{R}$ and $m$ are homogeneous in the interface layer, then energy spectrum is $\pm E_{\pm}=\pm\sqrt{p_{x}^2+p_{y}^{2}+(t_{R}\pm m)^2}$ as mentioned above. Now suppose the mass terms vary with position and that there is a region near the magnet where $|m|>|t_{R}|$ and a region past the extent of the magnet where $|m|<|t_{R}|.$ In this case, there exists a mass domain wall which necesarily binds propagating states along the 1D domain wall. Heuristically it is easy to see the character of the states that must be present: if instead of a domain wall between $m$ and $t_{R}$ we had a domain between $+t_R$ and $-t_{R}$, then the boundstates in this case would be a single-pair of time-reversed counter-propagating modes (\emph{i.e.} a helical metal) since this configuration is equivalent to an insertion of a $\pi$-flux tube\cite{wormhole}. In the case at hand, the $m/t_{R}$ domain-wall will only contain a single \emph{chiral} fermion since the effect of $m$ is to  generate a domain-wall for only one of the members of the Kramers' pair. Said a different way, in a homogenous system, if $m=0$ and we take $t_{R}$ from negative to positive then the critical point is time-reversal invariant and occurs where four-bands touch. If instead we take $m\neq 0$ and tune from $|t_R|<|m|$ to $|t_R|>|m|$, then the critical point happens only between two-bands and is essentially half of the time-reversal invariant phase transition. This means that only one of the Kramers' partners sees a domain wall. This is mathematically similar to the chiral superconductor phase transition presented in \cite{qi2010}. We can also analytically solve for the boundstate given that we have periodic boundary conditions in the y-direction and that the mass-domain wall occurs in the x-direction. For a domain wall where $m\neq 0$ for $x<0$ and $t_R \neq 0$ for $x>0$ we pick the ansatz
\[
\psi_{0}(p_y)=\xi_0 e^{ip_y y}\times\begin{cases}
\begin{array}{c}
e^{mx}\ \ \ x<0\\
e^{-t_{R}x}\ \ \ x>0
\end{array}\end{cases},
\]  where $\xi_0 $ is a constant spinor. We find the solution of the Dirac equation with $\xi_0=1/2(1,1,-1,1)^{T}$ and an anti-chiral dispersion relation $E(p_y)=- p_y$ as we expected.

Thus, we find that for a domain wall between a magnetic region and a tunneling dominated region, there exists a chiral interface state, giving rise to a quantum Hall effect. The conventional way to generate the quantum Hall effect via a magnetic domain wall on a single TI surface is topologically equivalent to our construction as seen in Fig. \ref{fig:3dfolding}. The equivalence can be understood by starting with a single TI with a magnetic domain wall on the surface and then deforming and folding the surface until it becomes a domain wall between a tunneling region and a single-domain magnetic region. 
Note that the folding picture works for any direction of the magnetization, assuming that a gap is opened at the surface, \emph{i.e.}, that the magnetization is not exactly parallel to the surface. The quantization of the Hall conductance can also be seen following the arguments of Ref. \cite{qi2008b} by integrating the magneto-electric polarizability $P_3$ around a loop enclosing the domain wall/hinge region. $P_3$ is well-defined since the system is gapped along the entire line of integration and we find 
\begin{equation}
\sigma^{2D}_{xy}=\frac{1}{2\pi}\frac{e^2}{\hbar}\oint dP_3=\frac{e^2}{h}\label{eq:P3int},
\end{equation}\noindent as expected for a single-chiral edge state. In calculating this we have used the fact that $P_3$ is odd under time reversal and thus must wind in opposite directions when passing from the 3D TI bulk through magnetic layers having opposite polarizations. Thus,  we have shown that TI structures with only \emph{one} magnet can generate an integer quantum Hall effect. 

To discuss the consequences of  magnet-tunneling competition for phenomena in the 2D TI (quantum spin Hall effect), as shown in Refs. \cite{qi2008,qi2008b}, in the presence of an anti-phase magnetic domain wall, a half-charge is localized on the edge of the 2D TI at the location of the domain wall. We demonstrate that on an $m^{++}-t_{R}$ domain wall there is also a localized half-charge. We note that the folding procedure in Fig. \ref{fig:3dfolding} still applies in 2D but with the chiral modes replaced with a bound $\pm e/2$ charge. We now provide a more general argument for the existence of the $e/2$ charge on a purely magnetic domain wall on a single edge than that presented in Ref. \cite{qi2008} using the bosonization formalism, which is thus incorporates non-vanishing interactions. We then carry out the argument for two edges with tunneling to show that indeed a half-charge is induced in that case as well. 
The propagating states of a single QSH helical edge state are described
by the Hamiltonian\cite{wu2006} 
\begin{eqnarray}
H_{0} & = & v\int dx\left[\psi^{\dagger}(x)p_x\sigma^z\psi(x)\right]
\end{eqnarray}\noindent where $\psi(x)=(\psi_{R\uparrow},\psi_{L\downarrow})^{T}.$ 
The coupling of the edge to a magnetic island can be described by 
\begin{equation}
H_{m}=-J\mu_{B}\int dx(m_{-}\psi^{\dagger}(x)\sigma^{+}\psi(x)+h.c.),\label{eq:magcoupling}
\end{equation} where $\mu_{B}$ is the Bohr magneton, $J$ is an exchange coupling
constant $\sigma^{\pm}=1/2(\sigma^x\pm i \sigma^y)$ and $m_{\pm}=M_{x}\pm iM_{y}=|m|e^{\pm i\theta_{H}}$ reflects the magnetization.
We can bosonize the Hamiltonian using $\psi_{R\uparrow}(x)\sim e^{-i(\phi(x)-\theta(x))},\ \psi_{L\downarrow}(x)\sim e^{i(\phi(x)+\theta(x))}$ to get
\begin{eqnarray}
H&=&\frac{1}{2\pi}\int dx\left[uK(\nabla\theta)^{2}+\frac{u}{K}(\nabla\phi)^{2}\right. \nonumber\\ &-&\left.\frac{2J\mu_{B}|m|}{\alpha}\cos(2\phi(x)-\theta_{H})\right],\label{eq:Bosonization}
\end{eqnarray}
From standard results\cite{GiamarchiBook}, the exchange coupling is relevant for $K<2$ (as we assume) and  at low temperatures, $\phi$ is locked to the energy minima
\begin{equation}
\phi(x)=n\pi+\frac{\theta_{H}}{2},\label{eq:phi frozern}
\end{equation}\noindent where $n$ is an integer. 
Now suppose we make $m_{+}(x)$ inhomogeneous with the domain wall profile $m_{+}(x)=|m|e^{i\theta_{H}^{L}}$ for $x<0$ and $m_{+}(x)=|m|e^{i\theta_{H}^{R}}$ for $x>0.$ This gives rise to the inhomogeneous form $\phi(x)=n\pi +\frac{\theta_{H}^L}{2}$ for $x<0$ and $\phi(x)=\ell\pi +\frac{\theta_{H}^R}{2}$ for $x>0$.
The charge density has the form $\rho(x)=-\frac{1}{\pi}\nabla\phi$ and thus the charge trapped on the domain wall is 
\begin{equation}
Q_{DW}=-\frac{1}{\pi}(\phi(x>0)-\phi(x<0))=(\ell-n)+\frac{1}{2\pi}\left(\theta^{L}_{H}-\theta^{R}_{H}\right).
\end{equation} For an anti-phase domain wall the formula gives $Q_{DW}=(n-\ell)+\frac{1}{2}$ as expected.

Turning to the two coupled 2D TI system,  the Hamiltonian for two corresponding edges is given by
\begin{eqnarray}
\mathcal{H}^{(2)}_{0} & = &v\int dx \Psi^{\dagger}(x) \left[\begin{array}{cc}
p\sigma_{z} & 0\\
0 & -p\sigma_{z}
\end{array}\right]\Psi(x),\label{eq:Hk4} 
\end{eqnarray}
where  $\Psi(x)=(\psi_{tR\uparrow}\;\; \psi_{tL\downarrow}\;\; \psi_{bR\downarrow}\;\;\psi_{bL\uparrow})^{T}$ and $t,b$ indicate top and bottom edges.  
To bosonize this Hamiltonian, care requires defining the fermions as 
$\psi_{tR\uparrow}(x)  =  \frac{U_{tR}}{\sqrt{2\pi\alpha}}e^{-i(\phi_{t}(x)-\theta_{t}(x))}$, $\psi_{tL\downarrow}(x)=\frac{U_{tL}}{\sqrt{2\pi\alpha}}e^{i(\phi_{t}(x)+\theta_{t}(x))},$ $\psi_{bR\downarrow}(x) = \frac{U_{bR}}{\sqrt{2\pi\alpha}}e^{-i(\phi_{b}(x)-\theta_{b}(x))},$  $\psi_{bL\uparrow}(x)=\frac{U_{bL}}{\sqrt{2\pi\alpha}}e^{i(\phi_{b}(x)+\theta_{b}(x))}$ where $\alpha$ is the momentum cut-off and the  $U_{ab}$ are the Klein factors preserving electron anti-commutation rules, where $U_{tR}^{\dagger}U^{\phantom{\dagger}}_{bL}U_{bR}^{\dagger}U^{\phantom{\dagger}}_{tL}=-U_{tR}^{\dagger}U^{\phantom{\dagger}}_{tL}U_{bR}^{\dagger}U^{\phantom{\dagger}}_{bL}$\cite{GiamarchiBook}.

Let us consider the perturbing mass terms for the double-edge system. The magnetic region couples to the two edges independently leading to two copies of Eq. \ref{eq:magcoupling} which when bosonized yields:
\begin{eqnarray}  
H^{(2)}_{m}&=& -\frac{J\mu_{B}|m|}{\pi\alpha}\int dx(\cos(2\phi_{t}(x)-\theta_{H})\nonumber\\
& &\ \ \ \ \ \ \ \ \ \ +\cos(2\phi_{b}(x)+\theta_{H}))
\end{eqnarray}\noindent where we used the choices $U_{tR}^{\dagger}U^{\phantom{\dagger}}_{tL}=U_{bR}^{\dagger}U^{\phantom{\dagger}}_{bL}=U_{tR}^{\dagger}U^{\phantom{\dagger}}_{bL}= -U_{tL}^{\dagger}U^{\phantom{\dagger}}_{bR}=1$ which satisfies the constraint above. 
The inter-edge tunnel coupling is 
\begin{eqnarray}
H_{t} & = & -t_{R}\int dx\ (\psi_{tR\uparrow}^{\dagger}\psi_{bL\uparrow}+\psi_{tL\downarrow}^{\dagger}\psi_{bR\downarrow}+h.c.)\nonumber\\
 & = &\frac{-2t_{R}}{\pi\alpha}\int dx(\sin(\phi_{t}+\phi_{b})\sin(\theta_{t}-\theta_{b}))
\end{eqnarray} Thus, since $J$ and $t_R$ are relevant for weak interactions ($2-\sqrt{3}<K<2$) we know that both terms will lock their phases in regions where they are present giving:
\begin{equation}
(\phi_{t}+\phi_{b})(x)\text{=}\begin{cases}
\begin{array}{c}
n\pi\ \ \ \ \ \ \ \ \ \text{magnetic region}\\
\frac{\pi}{2}+l\pi\ \ \ \ \ \ \text{tunneling region }
\end{array}\end{cases}
\end{equation}\noindent where $n,\ell$ are integers. 
The total charge density  is $\rho(x)=-\frac{1}{\pi}\nabla(\phi_{t}+\phi_{b})$
yielding a trapped charge  on a magnetic/tunneling domain wall:  $Q_{DW}=q+\frac{1}{2}$
 for an integer $q.$ We remark that the Klein factors, and thus Fermi statistics (since electrons can now exchange positions by moving to the other TI and then coming back), were crucial for this derivation; one only finds integer charge if they are neglected. A similar topological argument to Eq. \ref{eq:P3int} based on the topological electromagnetic response of 2D $Z_2$ topological insulators given in Ref. \cite{qi2008b} can also be given.

\begin{figure}[t]
    \begin{center}
        \includegraphics[width=3.5in]{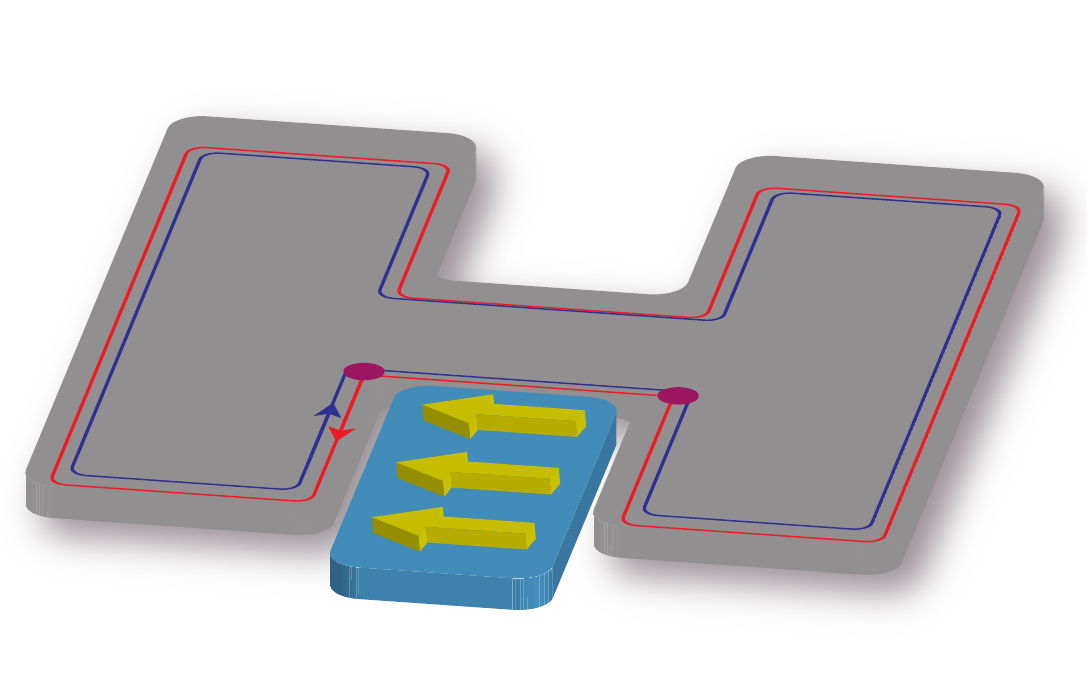}
    \end{center}
    \caption{Quantum spin Hall state in an H-bar geometry. Edge states conform to the geometry and in the lower half  travel around an inset ferromagnet whose magnetization direction is shown by the arrows. An $e/2$ charge is localized between the two spots at the corners of the ferromagnet. In the ideal case with this geometry there are $e/4$ charges localized at the location of each corner spot.}
    \label{fig:hbar}
\end{figure}

The folding picture provides a useful illustration of the tunneling domain, but there are some other geometries where the same physics is also apparent.  Notably one can study the ``H-bar" geometry, such as that used in non-local transport experiments in HgTe/CdTe quantum wells\cite{roth2009}. In Fig. \ref{fig:hbar} we have shown such a geometry flanked by a ferromagnet in one of the U-shaped regions of the `H.' The presence of a magnet-tunneling domain wall, which should give rise to $e/2$ charge localized on the cross-bar, can be seen by treating the two vertical legs as two separate QSH systems connected by a small strip of tunneling (the crossbar). Physically, the geometry replaces the folding.  The edge state, upon entering  the first part of the `U'  from the left,  experiences a magnetization that points to the left relative to it's velocity while, upon exiting the `U' on the right, experiences a relative magnetization pointing to the right. Thus, skirting around the magnet yields a changing magnetization, giving rise to the localized charge. If the path does not perfectly reverse direction, the charge is not quantized to be perfectly $e/2$ since the edge electron does not encounter a crisp, anti-phase domain-wall. The charge is given by the relative angle between the incident and exiting effective magnetizations in units of $e/2\pi$ and, for instance, could split into two $e/4$ charges localized near the corners where the relative magnetization typically changes by $\pi/2.$ An extension of this geometry to the 3D case would yield a chiral edge state. It is important to note in that case that the existence of the chiral modes \emph{does not} depend on the precise reversal of the path direction.


To summarize, we have shown that two sought after phenomena-- quantum Hall states and fractional charge,-- can be achieved in 2D and 3D TI's with experimentally viable fabrication. In 3D one must simply grow a magnetic layer sandwiched between two TI layers, and in 2D one must simply fabricate an H-bar geometry and deposit a magnetic island on one of the indentations of the H. We also performed a cursory examination of similar geometries where the magnet was replaced by an s-wave superconductor and found that the effects, while interesting, were not feasible for current experimental capabilities.

\begin{acknowledgments}
QM would like to thank J. Teo for useful discussions. All authors of this work are supported by the U.S. Department of Energy, Division of Materials Sciences under Award No. DE-FG02-07ER46453.
\end{acknowledgments}

%

\end{document}